\documentclass{llncs}
\usepackage{verbatim}
\usepackage{amsmath}
\usepackage{stmaryrd}
\usepackage{color}
\usepackage{hyperref}
\usepackage{amsmath, amssymb}
\usepackage{url, listings}
\usepackage{xspace}

\usepackage{listings}
\definecolor{vccOutColor}{rgb}{0.0,0.0,0.3}
\definecolor{vccDotsColor}{rgb}{0.3,0.3,0.3}
\lstdefinelanguage{VCC}[ANSI]{C}{%
  morekeywords={axiom,requires,ensures,reads,writes,def,maintains,pure,datatype,read_only,inv,inv2,this,full_extent,extent_mutable,
                assert,assume,invariant,unwrapping,wrap,unwrap,atomic,dynamic_owns,wrapped, mutable,natural,extent,thread_local_array,bool,
		unchecked,ghost,atomic_inline,atomic_read,claim,make_claim,old,unchanged,destroy_claim,approves,owns,owner,claimable,mine,
		__vcc_spec,logic,always,bump_volatile_version,ghost_atomic,on_unwrap,true, false,
		},
  moretexcs={proper,
                thread,me,closed,consistent,span,
		result,any,forall,exists,lambda,in,diff,inter,set,domain,
                union_with,diff_with,fresh,
                claim_struct,claims_object,claims,
		at,active_claim,state,},
  moredelim=[is][\fontfamily{cmtt}\selectfont\textcolor{vccOutColor}]{/*`}{`*/},
  morecomment=[is]{//--}{//--},
}[tex]

\lstnewenvironment{VCC}[1][]{%
  \lstset{%
    floatplacement={tbp},
    captionpos=b,language=VCC,#1
    }}{}%

\lstnewenvironment{VCC2}[1][0.45]
    {\global\let\lst@intname\@empty
     \gdef\lst@sample{}%
    \setbox0=\hbox\bgroup
         \setkeys{lst}{floatplacement={tbp},captionpos=b,language=VCC}
         }
    {\egroup%
      \noindent\begin{minipage}{#1\linewidth}\hbox to\linewidth{\box0\hss}\end{minipage}}

\def\vcc{%
    \lstinline[language=VCC,keywordstyle=]}

\usepackage[T1]{fontenc}
\usepackage{times}

\newcommand{\secref}[1]{\hyperref[sect:#1]{\textsection~\ref{sect:#1}}}

\newcommand{\Def}[1]{\textit{\textbf{#1}}}

\title{Modular Verification of Hybrid System Code with VCC}
\author{Ernie Cohen (ernie.cohen@acm.org)}
\institute{Department of Computer and Information Sciences, University of Pennsylvania}

\begin{document}
\maketitle 
\begin{abstract}
We present a methodology for object-modular reasoning about hybrid system code using VCC, a deductive verifier for concurrent C code. We define in VCC an explicit time model, in which the passage of time must respect the invariants of certain timed objects. Fields that change automatically with changes to time are then defined as volatile fields with suitable invariants. We also define two types of timed objects that prevent time from advancing past a given expiration: Timers (which represent assumptions about the upper limit on the time it takes to do something) and Deadlines (which represent assertions about these limits). The difference between the two is that once the expiration time of a Deadline is reached, the Deadline and time itself are permanently deadlocked. Our methodology includes showing that all Deadlines are eventually destroyed, proving that they do not interfere with the flow of time.

Untimed objects can (transitively) own (or otherwise depend on) timed objects, and use their timing behavior to justify the admissibility of object invariants that depend on time not moving too far forward. This allows us to reason about timed and hybrid system code within the object-modular methodology of VCC.

As an example, we verify the code of a toy device driver for a toy steam boiler. 
\end{abstract}

\thispagestyle{empty}
\pagestyle{empty}

\lstset{
  basicstyle=\small\ttfamily,
}

\lstset{
  keywordstyle=\bfseries, 
  breaklines=true,
  breakatwhitespace=true,
  numberstyle=\tiny\sf,
  escapeinside={/*-}{*/},
  numbers=none,
  emptylines=1,
  rangeprefix=\/\*\{,
  rangesuffix=\}\*\/,
  includerangemarker=false, 
}

\section{Object-Modular Reasoning About Timed and Hybrid Systems}
In recent years, thanks to advances in automatic deduction (e.g. SMT) and verification methodology, 
deductive functional verification of real-world C code has become increasingly practical, using verifiers such as VCC \cite{VCC}\cite{Admissibility} and Verifast \cite{Verifast}.
While such verification is expensive compared to pushbutton technologies like static analysis or model checking, it's ability to handle fine-grained concurrency, along with its potential to reduce testing and maintainance costs while provide superior internal documentation, is making it commercially viable, even for COTS software.

One of the methodological developments that has made this productivity possible has been the use of what might be called ``object-modular'' reasoning. The hallmarks of this reasoning are (1) the state is partitioned into a number of objects, each with its own invariant (which might talk about the entire state); (2) verifying the safety of an update to the state requires checking
only the invariants of the objects that are updated; and (3) reasoning about the code itself is sequential. An essential ingredient of
object-modular reasoning is that the object invariants are ``admissible'', or ``self framing'', which guarantees that the invariant of an object is not broken by updates to other objects that preserve the invariants of the updated objects.

One of the most important applications of C is to write software that interacts with devices (and/or the outside world, which is modeled similarly). 
In object-modular verifiers, devices are conveniently understood as separate threads.
(Alternatively, in a tool like VCC that provides 2-state object invariants, devices can be modeled directly as objects.)
This approach has proven practical for verifying safety properties of asynchronous hardware-software systems, such as hypervisors (where the MMU has to be considered a concurrent co-processor\cite{TLB}). 
A natural way to extend this approach to timed systems is to model time as an explicit variable that is updated by separate discrete actions \cite{LamportReal}.
We can take advantage of time in two different ways. First, we can capture the times at which certain program points are reached by recording the time in a ghost variable, and 
express assumptions about the timing behavior of a piece of code by making assumptions on these variables.
Second, we can identify certain objects as ``timed'', and require that changes to time to respect the invariants of these objects.
One can then reason about the combined program using ordinary invariance reasoning.
This approach can be extended to handle rely-guarantee style composition \cite{LamportReal}.
To handle hybrid systems, we can add explicit state variables for every observable that changes autonomously with time, requiring these variables to be updated appropriately when time moves forward.

In this paper, we propose a methodology for object-modular explicit-time verification. 
It is built on top of the methodology of VCC, namely admissible 2-state invariants \cite{Admissibility}.
This underlying machinery makes it easy to encode time and timed objects, because the object-modular structure of VCC allows a collection of objects to be viewed 
simultaneously as a conjunction (of their invariants) and as a disjunction (of their transition relations).
It also makes it easy to uniformly treat those objects representing assumptions about time (timed objects) and objects representing invariants about time.
We implement the methodology on top of VCC, defining C and timed objects using ordinary VCC types. 
The result is a system in which we can  directly verify C code that interacts with a timed environment, without having to go through an intermediate model.

We illustrate the methodology with the verification of a toy driver for a toy steam boiler.


\section{Methodology Overview}
In VCC, the environment can perform, at any time, any change to the state that respects the invariants of all updated objects. 
We take advantage of this by putting time in an object and constraining the update of time with invariants, rather than trying to write down explicit updates to time. Thus, time is updated by the environment.
We identify certain objects as \Def{timed}. Being timed simply means that the invariant of the object must be respected by state changes that update time.
In general, the invariants of timed objects may require parts of the state to be updated in the same step in which time advances.
Thus, as time moves forward, many object fields may change simultaneously.

In a typical hybrid control program, correctness depends on certain things being done within specified time bounds. 
For example, the safety of a control system requires a driver to periodically check sensors and adjust the controls accordingly to keep the system in a good state.
Without sufficiently alert response, a failure can result from time being moved too far forward (perhaps in several steps), accompanied by changes to fields that change with time.
Thus, we can rephrase the responsiveness requirement as an invariant that keeps time from moving too far forward (i.e., beyond the point where a suitable control correction can be applied). 
We define two kinds of timed objects that can stop time from moving forward, \vcc{Timers} and \vcc{Deadlines}.

A \vcc{Timer} is a timed object with an expiration time and an invariant that time cannot move beyond expiration.
In order for a timer to not to eliminate executions in which time flows freely, the timer must be destroyed or its expiration time reset before the expiration time is reached.
Thus, a \vcc{Timer} corresponds to an assumption that its expiration is never exceeded (while it exists) in actual execution.

We could define an assertional (as opposed to assumptional) version of a \vcc{Timer} (say, an \vcc{AssertionalTimer}) by defining a type with the same invariant but making it untimed.
We could make its invariant admissible by having it own a Timer whose expiration is no later than its own. However, the resulting object would be no more useful for reasoning than using the \vcc{Timer} itself.

Instead, we define another kind of ``assertional timer'', as follows. A \vcc{Deadline} is just like a \vcc{Timer}, except that it has two additional invariants that say that once time reaches the expiration of the \vcc{Deadline},
(1) the deadline can never be destroyed, and (2) the deadline expiration time cannot be changed. 
Thus, once a \vcc{Deadline} expires, both it and time itself are forever locked into a deadly embrace. 

\vcc{Deadline}s would obviously be unsound (as assertions) and useless without further refinement; we could simply create a \vcc{Deadline} that has expired and use it to deadlock time forever.
The twist is that we require (as part of verification) proof that each \vcc{Deadline} is destroyed.
If this proof succeeds, we know that no \vcc{Deadline} ever reaches expiration,
which means that \vcc{Deadline}s cannot interfere with the flow of time.
More formally, destruction of all \vcc{Deadlines} means that the program with \vcc{Deadline}s simulates the program without \vcc{Deadline}s.

The advantage of a \vcc{Deadline} over an \vcc{AssertionalTimer} or a \vcc{Timer} is that \vcc{Deadline}s are immediately admissible, without them having to directly reference \vcc{Timer}s or other \vcc{Deadline}s. 
A \vcc{Deadline} can be justified using reasoning within the context of the thread that destroys it, without having to make that justification part of the \vcc{Deadline} structure itself. Thus, the use of \vcc{Deadline}s provides a clean separation between
the users of the \vcc{Deadline} (who use its invariant) and the provider (the thread that guarantees that it will be pushed forward).

Since destroying a \vcc{Deadline} is a progress property, rather than a safety property, we prove that \vcc{Deadline}s are destroyed by allowing \vcc{Deadline}s to be created only on the (ghost) stack of code that is proved to terminate.
This proof of termination can make use of object invariants, including invariants with \vcc{Deadline}s - perhaps even \vcc{Deadline}s created by the same thread.
This does not create cyclical reasoning, because destroying the \vcc{Deadline} requires destroying it strictly before its expiration, whereas the same \vcc{Deadline} is useful only to prove that time doesn't move beyond its expiration.

\section{VCC Background}

VCC is a deductive verifier for concurrent C code, developed at Microsoft. VCC has been used to verify over 100 KLOC of concurrent product code, mostly from low-level software such as hypervisors, kernels, OLTP, lock-free data structures and so on. Verification is modular wrt. threads, objects, and functions. We give here only an outline of the VCC methodology; more details can be found on the VCC homepage (\vcc{vcc.codeplex.com}).

In VCC, a program operates on a fixed set of typed objects; the identity of an object is given by its address and its type. (For example, for
each user-defined \vcc{struct} type there is an object of that type for each address properly aligned for objects of that type.) Each object has
a collection of fields determined by its type. A state of the program is a function from objects and field names to values. One field of
each object is a Boolean indicating whether the object is valid; the valid objects represent those objects that actually exist in a given
state, so object creation/destruction corresponds to object validation/invalidation.

A transition is an ordered pair of states, a prestate and a poststate. Each object has a two-state invariant (a predicate on
transitions); these invariants can mention arbitrary parts of the state, and are mostly generated directly from the annotations on the
program. The invariant of a type/object is the conjunction of all invariants mandated in this manual. An invariant can be interpreted as
an invariant on a single state by applying it to the transition from the state to itself (stuttering transition). A state is good if all object
invariants hold in that state; a transition is good if it satisfies the invariant of each object. A sequence of states (finite or infinite) is good
iff all of its states are good and the transitions between successive states of the sequence are good.

A transition updates an object if some field of the object differs between the prestate and poststate of the transition. A transition is
legal if its prestate is not good or if it satisfies the invariants of all updated objects. A sequence of states is an legal execution iff the
initial state is good and the transition from each nonterminal state of the sequence to its successor is legal. The program is good iff every
legal execution of the program is good; successful verification of the program shows that it is good.
A type is admissible iff, for every object o of the type, (1) every legal transition with a good prestate satisfies the invariant of o,
and (2) the poststate of every good transition from a good state satisfies the invariant of o. It is easy to prove by induction on legal
executions that if all object types are admissible, the program is good. The program is verified by proving that every type of the program
is admissible. (There is a type corresponding to each function; admissibility if this type shows that changing the state of the system
according to the operational semantics of the function is legal.)

Each valid object has a Boolean ghost field that says whether it is closed, and a ghost field that gives its owner (which is also an
object). User-defined object invariants are guaranteed to hold only when the object is closed. Threads are also modeled as objects; in
the context of a thread, an object owned by the thread is said to be wrapped if it is closed, and mutable if it is open (not closed). Only
threads can own open objects. The owner of an open object has ``exclusive'' use of the object, and so can operate on it sequentially. An
implicit invariant is that nonvolatile fields of objects don't change while the object is closed, so such fields can also be read sequentially if
the object is known to be closed. Fields of closed objects can be updated only if they are marked as volatile; such fields can be accessed
only within explicitly marked atomic actions.

Each object field is either concrete or ghost; concrete fields correspond to data present in a running program. Each concrete field
of each object has a fixed address and size. Each object has an invariant saying that concrete fields of valid objects don't overlap, and
verification of a function is that it accesses only fields of valid objects; thus, every legal execution of good program is simulated by a
legal execution in which concrete field accesses are replaced by memory accesses to shared C heap. The program can include ghost code
not included in the unannotated program; however, verification guarantees that all such code terminates, and its execution does not
change the concrete state. This implies that every legal execution of the unannotated program is the projection of a legal execution of
the program, which allows properties of the program to be projected to properties of the unannotated program.

Each thread has a field that gives its ``local copy'' of the global state. When a thread updates an object, it also updates the local copy;
when it reads an object, it reads from the local copy. Just before each atomic update, the thread updates its local copy to the global
state. It is an invariant of each thread that for any object that it owns, the local copy agrees with the actual object on all fields if the
object is open and on all nonvolatile fields if the object is closed. However, assertions that mention objects not directly readable by the
thread (without using an atomic action) might reference fields where the local and global copies disagree; thus, assertions that appear in
the annotation of a thread are guaranteed to correspond to global assertions only at the beginning of an atomic action. This machinery
allows us to ``pretend'' that threads are interrupted by other threads only just before entering explicit atomic actions.

\section{Time and Timed Objects}
We present our treatment of hybrid systems with the VCC code used to verify a toy example. We begin with the representation of time.

\begin{VCC}
_(struct Time {
\end{VCC}
VCC annotations are enclosed in the macro \vcc{_(...)}, which causes annotations to be ignored by the C compiler. In this case, it means that the type being declared
is a ghost type. 

\begin{VCC}
    volatile \integer cur;
\end{VCC}
We represent the current time as a field \vcc{cur} of type \vcc{\integer}, a built-in VCC type representing mathematical integers. 
(All built-in VCC keywords that can appear in code begin with a backslash, to avoid collision with user-defined and C identifiers.)
Although we defined time to be an integer, we could as well have declared it to be real or rational. 
The field \vcc{cur} is declared as \vcc{volatile}, a keyword which in C carries the meaning that its value might be changed by forces outside of the program text itself. 
To VCC, being a volatile field means that \vcc{cur} can change while the object of which it is a field is closed. 
 \begin{VCC}
    _(invariant  \old(cur) <= cur)
\end{VCC}
When writing object invariants, \vcc{\old(e)} means the value that the expression \vcc{e}
had in the prestate of a transition, while expressions not surrounded by \vcc{cur} are evaluated in the poststate of the transition.
This invariant says that for any object \vcc{o} of type \vcc{Time}, \vcc{o->cur} can only increase; this expresses the notion that time can only move forward.
\begin{VCC}
    volatile \bool timed[\object];
\end{VCC}
Associated with an object of type \vcc{Time} is a set of timed objects. VCC uses C array syntax to define maps, with the domain of the map in place of the size of the array. 
The declaration above defines \vcc{timed} as a map from \vcc{\object} (a built-in VCC type representing all typed pointers) to Booleans. 
This map is a primitive value; again, it is declared as \vcc{volatile} because the set of timed objects can change during execution.
Note that unlike C arrays, a map represent a single, primitive value; you can take the address of the map, but not the address of a component of the map.

\begin{VCC}
    _(invariant \forall \object o; 
        \old(timed[o]) ==> timed[o] || \inv2(o))
\end{VCC}
The set of timed objects can change; this invariant says that when an object \vcc{o} goes from being timed to being untimed, the state change must satisfy the 
2-state invariant of \vcc{o}. This invariant is needed in order to make admissible invariants of objects that require them to be in the set of timed objects.
More colloquially, it prevents  actions from simply removing an object from the set without the object's permission.
\begin{VCC}
    _(invariant \forall \object o; timed[o] && o->\closed 
            ==>  \unchanged(cur) ||  \inv2(o))
\end{VCC}
The last invariant of \vcc{Time} says that when time changes, it does so in a way that respects the invariants of all timed object. 
In particular, a change to the time has to change any fields of timed objects that are required (through the invariants of the timed object) to change with time.
Note that we do not define any specific action for changing time (since such an action would need to know the invariants of all of the timed objects);
instead, we think of time as (possibly) moved forward by some force in the environment (e.g., God).
\begin{VCC}
} time;)
\end{VCC}
We introduce a single object \vcc{time} of type \vcc{Time}.
\begin{VCC}
_(axiom (&time)->\closed);
\end{VCC}
For simplicity, we assume that the object \vcc{time} is eternal, i.e. always closed. Without this, there would be nothing to stop God from coming in opening up time
and moving time backward.

\begin{VCC}
// timed objects are listed in time.timed
#define TIMED \
    _(invariant  \this->\closed ==> time.timed[\this]) \
    _(invariant (&time)->\closed && \inv2(&time))
\end{VCC}
Being a timed object is essentially a "mixin" that we add to a type, declaring that type to be timed. 
Note that ordinary programmer-defined types should never be timed; making a type timed essentially makes an assumption about its behavior when time changes.
Declaring type as \vcc{TIMED} just adds the above object two invariants. The first says that when the object
is closed, it is in the list of timed objects. (This invariant is admissible only because of the lst invariant of type \vcc{Time}.)
The second invariant is not logically required, but simply allows VCC to note when
reasoning about such an object that \vcc{time} is closed and that its invariant is relevant. 
This can be considered essentially a hint for VCC.

Normally all timed types are ghost; an exception is that when modeling hardware, where memory-mapped
devices that react to time would normally be modeled as concrete data, since otherwise reading such a device
would require information flow from ghost state to concrete state, which would cause VCC to complain (since such
flow jeopardizes the soundness of using ghost code in general).

\begin{VCC}
// T == the current time
#define T (time.cur)
// dT == how far time just moved forward (atomically)
#define dT (time.cur - \old(time.cur))
\end{VCC}
For convenience, we define abbreviations for the current time, and, in the context of an environmental state transition, 
how far time is advancing in the transition. The latter is useful only when writing invariants of timed objects.

\section{Deadlines}

\begin{VCC}
_(typedef struct Deadline {
    TIMED
    // the expiration time
    volatile \integer t;
    _(invariant \approves(\this->\owner,t))  
\end{VCC}
This declaration follows the usual C idiom of defining a type within a \vcc{typedef}.
A \vcc{Deadline} is a timed ghost object with an expiration time. 
This expiration time is volatile, so that it can be reset without having to open up the \vcc{Deadline}.
Changes to the expiration time are ``approved'' by the owner of the \vcc{Deadline}.
This essentially means that (1) if the owner of a \vcc{Deadline} is a (non-thread) object, then on any change to the expiration time, the invariant of the owner must be satisfied, and
(2) if the owner of a \vcc{Deadline} is a thread, the field \vcc{t} cannot be changed by other threads (or the environment), and it is treated like a local variable for the purposes of framing.
(We'll see an example of this later.)

\begin{VCC}
    _(invariant T <= t)
    _(invariant \unchanged(t) || \old(T < t))
    _(invariant \on_unwrap(\this, \old(T < t)))
} Deadline;)
\end{VCC}
The first invariant says that time is prevented from moving past the expiration time. Note that this invariant is evaluated in the poststate of a transition, so in principle 
the environment could change time while simultaneously changing the expiration time, if the \vcc{Deadline} was owned by an object whose invariant allowed such a change.
(In our examples, this cannot happen because we explicitly state who controls the expiration times of the \vcc{Deadline}s we use.) 
The second invariant says that the expiration time cannot be changed once \vcc{Deadline} expires.
The third invariant says that while the  \vcc{Deadline} is expired, it can never be destroyed. Since time cannot move backwards, once a \vcc{Deadline} expires, it can never unexpire, so 
both the \vcc{Deadline} and time itself are frozen forever. 

Conversely, if a \vcc{Deadline} is destroyed, it is guaranteed to never have been missed.

\begin{VCC}
_(void DeadlineNew(Deadline ^d, \natural delta)
    _(writes \extent(d))
    _(ensures \wrapped(d) && d->t - T == delta)
    _(ensures \unchanged(T))
\end{VCC}
The function for initializing a \vcc{Deadline} writes the \vcc{\extent} of the \vcc{Deadline} (which includes all of its fields); this implicitly requires that the \vcc{Deadline} is mutable
(i.e., open and owned by the thread executing the code).
The type \vcc{Deadline ^} is the type of pointers to ghost objects of type \vcc{Deadline}, analogous to the ordinary C type \vcc{Deadline *}.  
The postconditions of the function are that the \vcc{Deadline} is ``wrapped'' (i.e., closed and owned by the current thread), with an expiration time \vcc{delta} units ahead of the current time.

The last postcondition says that no time passes between when the function is called and when it returns. This is not surprising for ghost code; VCC semantics guarantees
that thread switches do not occur during execution of ghost code, so time could only be advanced in the body of a ghost function only if the function explicitly updated
the time (which it is allowed to do). 

While it is not made explicit in the function contract, every ghost function is required to terminate. This is because the soundness of using ghost code to reason about programs
presupposes that the ghost code terminates (so that the program running with the ghost code can be proved to simulate the program running without the ghost code). 
\begin{VCC}
{
    d->t = T + delta;
    _(ghost_atomic &time {
        time.timed[d] = \true;
    })
    _(wrap d)
})
\end{VCC}
The \vcc{Deadline} expiration is set using the current time. Note that ghost code can atomically read any part of the state, even if that state is volatile, without the code reading that state being inside an atomic action.
However, the update to the set of timed objects to include the new \vcc{Deadline} does have to be inside an atomic action, so as to invoke a check on the invariant of \vcc{time}.
The invariants of the \vcc{Deadline} having been establish, we wrap the \vcc{Deadline} (which requires a check of the invariant of that object).

An important part of the methodology for \vcc{Deadline}s is to verify that each \vcc{Deadline} is eventually destroyed. 
We achieve this by allowing \vcc{Deadline}s to be allocated only on the stack, and only inside the scope of a function or loop that is proved to terminate. 
Note that VCC currently has no way to enforce this restriction, though it could be easily added.

\section{Timers}
There are two ways to handle lower bounds on execution time. 
One possibility is to define them in essentially the same way as \vcc{Deadline}s, except leaving out the invariant that prevents \vcc{Timer}s from being destroyed if time ever reaches the expiration time:
\begin{VCC}
_(typedef struct Timer {
	TIMED
	volatile \integer t;
	_(invariant T <= t)
	_(invariant \approves(\this->\owner,t))
} Timer, ^PTimer;)
\end{VCC}
An advantage of this approach is that it makes it easy to control the use of \vcc{Timer}s, e.g. by allowing them only to bound the time to execute within a block. 
Another advantage is that it keeps time from exceeding the expiration of the \vcc{Timer} for the entire time from the timer creation until it is destroyed or its deadline is reset, rather than when it is explicitly checked.
Finally, \vcc{Timer}s can be used directly in object invariants.

\vcc{Timer}s have two disadvantages. 
First, if we create a timer and simply forget to reset it, the implicit assumption that results is not easily visible to the programmer.
Thus, \vcc{Timer}s have to be used in a disciplined way to avoid making unintentional assumptions.
Second, the expiration time of the \vcc{Timer} has to be chosen when the \vcc{Timer} is created, even though the desired \vcc{Deadline} might be dynamically determined by data available only later in the computation,
requiring the expiration time to be reset appropriately.
(Note, however, that this disadvantage doesn't occur in the current example.)

A simpler, and perhaps safer, way to handle upper bounds on execution time is to simply read the time, and later assume that not too much time has passed. The justification for using these assumptions directly is that a timer truly is an assumption about the timing behavior of a piece of code, so it is appropriate that checking that a timer hasn't expired should appear as an explicit assumption. 
For notational convenience, we ``create'' a new timer by simply declaring a ghost variable and setting it to the current time. We ``read'' that timer by just measuring how much time has passed since its creation.
\begin{VCC}
#define TIMER(name) \
    _(ghost \integer name) \
    _(ghost name = T) 

#define READ_TIMER(name) (T - name)
\end{VCC}
We can then prevent time from moving too far by simply assuming that the result of reading the timer is suitably small. Note that nothing prevents such an assumption from being inconsistent. 
We imagine that in a serious timed or hybrid verification, there would be intrinsics or other operations that have known time bounds, and that all time bounds would be limited to such intrinsics.

In some protocols, it is necessary to wait a minimum amount of time. This could be implemented using an assumption about the recorded times. As an alternative, it can be implemented using a loop that busy-waits for time to pass, performing a (non-ghost) atomic action each time through the loop to allow time to pass. Note, however, that such a loop could not be proved to terminate (since its termination depends on actions of the environment to actually push time forward). 

\section{Example: A Toy Steam Boiler}
We now give a simple toy example, a steam boiler. (We emphasize this as a toy example, and is not supposed to provide the depth of challenge of the original steam boiler specification problem \cite{SteamBoiler}; we use this example only because it is a familiar one.

\subsection{The Boiler Device}
\begin{VCC}
typedef struct Boiler {
    TIMED
    volatile int level;
    volatile BOOL on;
    _(invariant level == \old(level) + (\old(on) ? (int) dT : 0-(int) dT))
    _(invariant \approves(\this->\owner,on))
} Boiler; 
\end{VCC}
A boiler is a timed device with two variables: a continuously varying water level, and a boolean switch that turns the water on and off. 
The switch is controlled by the owner of the boiler, while the level changes only when time passes.
This example shows the typical specification form of like devices. The control component (the switch) is owner-approved, while the continuously varying part of the state is not.
An invariant controls how the continuously varying part of the state is required to change with time; in this case, the water level increases linearly when the water is turned on, and decreases linearly when the level is turned off. 
For simplicity, we do not include behavior beyond the reasonable physical boundaries, but it should be noted that in making this simplification we have somewhat cheated, in that the boiler above could actually be sued to eventually stop time (because of the bounded range of the type \vcc{int}).

Note that in making the water level an explicit concrete field, we allow the water level to be determined by code simply reading this field. This model would be appropriate when this capability is actually provided via a memory-mapped device. Alternatively, we could have made the water level a (volatile) ghost field, and allow it to be read only using some library function specific to the device.

\subsection{The Control Protocol}
Although it is not made explicit in the above definition of the device, our goal will be to control the water level of the boiler so as to keep it between 30 and 70.
We want to do this via a protocol that periodically checks the water level, and turns the water on if the level is below 50 and turns the water off if the level is above 50.
We capture the intent of this protocol with a ghost object as follows:
\begin{VCC}
_(typedef struct BoilerCtrl {
    Boiler *b;
    _(invariant \mine(b))  
\end{VCC}
Our boiler control protocol object is associated with a particular boiler object. The invariant above says that the protocol actually owns the boiler object, which really just means that the policy by which the water is shut on an off is controlled by this policy. (This ownership, or something like it, is needed to make the following invariants admissible.)
\begin{VCC}
    Deadline ^d; 
    _(invariant \mine(d))
\end{VCC}
The control policy also owns a \vcc{Deadline}. In this example, where the boiler is controlled by a single thread, we can make the \vcc{Deadline} a fixed object. We can be somewhat more flexible by making the pointer to the \vcc{Deadline} volatile, and making this pointer itself owner-approved. 

In any case, the use of the \vcc{Deadline} as part of the object is the key idea in the methodology. This \vcc{Deadline} prevents time from moving so far forward that the water might overflow or underflow.
\begin{VCC}
    volatile \integer deadline;
    _(invariant deadline == d->t)
    _(invariant \approves(\this->\owner,deadline))
\end{VCC}
This part of the protocol is to overcome a current limitation of approval in VCC. While the protocol is going to control the \vcc{Deadline}, we want the owner of the controller to also approve changes to the expiration time of the \vcc{Deadline}. However, VCC does not allow general owner approval (extending to threads) of arbitrary expressions, only fields of the object in which the approval invariant appears. 
Therefore, to allow the owner of the protocol to control updates to the \vcc{Deadline} expiration, we add an extra ghost copy of the expiration time to the protocol, and add invariants that this expiration is in sync with the expiration of the \vcc{Deadline}, and that this new shadow copy is owner-approved. (A more general approval mechanism should be available in the next version of VCC.)
\begin{VCC}
    _(invariant b->level <= 70 && b->level >= 30)
\end{VCC}
Here we see the invariant that the control policy is intended to maintain. If we included only this invariant, the control policy would fail the VCC admissibility check, because 
the passage of time, accompanied by the corresponding change to the boiler water level, could cause this invariant to be violated.
To prevent this from happening, we need some invariants that prevent time from moving far enough forward for this to happen. 
We have just such a mechanism at hand, namely the \vcc{Deadline}. The amount of time that we can afford to pass depends on the water level; hence the following invariants:
\begin{VCC}
    _(invariant b->on ==> b->level + d->t - T <= 70) 
    _(invariant !b->on ==> b->level - d->t + T >= 30)
} BoilerCtrl;)
\end{VCC}
Here we arrive at the crux of the matter. In order to make
These invariants say that the \vcc{Deadline} expiration is close enough to prevent time from moving forward far enough to break the previous invariant. 
Note that these invariants are preserved by the passage of time (thanks to the invariants of the boiler).

The interesting thing about the boiler controller is that it makes essential use of the invariant of the \vcc{Deadline}, even though the \vcc{Deadline} really only restricts the passage of time when 
its \vcc{Deadline} is reached, and such an occurrence would already be disastrous. Indeed, the  verification will show that  \vcc{Deadline}s in fact have no effect at all on the computation.
They can thus be viewed as ghost objects even relative to the other ghost objects, i.e. ghost objects that exist only to simplify the reasoning about the remaining ghost code.

\subsection{The Device Driver}

We now verify a concrete C program that implements the control policy of the last section. 
The basic idea is to set up a boiler control object that will own the boiler, and to provide a suitable \vcc{Deadline} for the controller.
The only additional responsibility of the thread is to prevent this \vcc{Deadline} from expiring. To prevent this expiration,
the driver periodically refreshes the \vcc{Deadline} expiration, turning the water on or off so as to preserve the invariant of the controller.

\begin{VCC}
void boilerDriver(Boiler *b)
    _(maintains \wrapped(b))
    _(writes b)
    _(decreases 0)
\end{VCC}
Before, we said that \vcc{Deadline}s can only be created on the stack of a block or function that is proved to terminate. 
Currently, VCC allows termination to be declared for a loop only if the surrounding function also terminates. 
Therefore, we include an explicit specification that the driver terminates. Of course, typical device drivers are not supposed to terminate (at least until the system crashes or is shut down), so this driver instead runs for a very, very long time. In essence, we are not so much proving termination as proving the possibility of termination.
\begin{VCC}
{
    _(ghost Deadline ctrlDeadline)
    _(ghost BoilerCtrl ctrl)
    _(ghost {
        ctrl.d = &ctrlDeadline;
        ctrl.b = b;
        DeadlineNew(&ctrlDeadline, 15);
        ctrl.deadline = T+15;
        (&ctrl)->\owns = {ctrl.d, b};
        _(assume b->level >=45 && b->level <= 55)
        _(wrap &ctrl)
    })
\end{VCC}
We begin by creating suitable \vcc{Deadline} and protocol objects, and setting up the controller to own the \vcc{Deadline} and the boiler.
We need that the boiler level to be suitably initialized to a level close to the target level; in a real implementation, this would be implemented using some sort of boilder initializaiton instead of an assumption.
Finally, we wrap up the control object; this involves a check of its invariants.

We then come to the main control loop.
\begin{VCC}
    for (unsigned i = 0; i < 10000000; i++)
        _(writes &ctrl)
        _(invariant ctrl.d == &ctrlDeadline)
        _(invariant \wrapped(&ctrl) && ctrl.d->delta > 10 && b==ctrl.b)
\end{VCC}
The control loop executes for many iterations. 
It's usually a good idea to give an explicit writes clause to a loop; in this case, we have to declare that we write \vcc{&ctrl} because we write to one of its owner-approved fields (namely \vcc{ctrl.deadline}).
Because we write this object, we need to speciy any invariants of the volatile fields of that object that need to be preserved by the loop, including that the object is wrapped.
Finally, whenever we reach the top of the loop, we have at least 10 time units left before the controller \vcc{Deadline} expires.
\begin{VCC}
    {
        TIMER(t1);
        _(assert &ctrl \in \domain(&ctrl) && ctrl.d \in \domain(&ctrl))
        _(atomic &ctrl, ctrl.b, ctrl.d) {
            _(assume READ_TIMER(t1) < 5)
\end{VCC}
The platform must guarantee that the following atomic takes place 
not too long after the top of the loop. To express this, we start
a timer at the top of the loop and assume no more than 5 timeunits have passed
when we get into the atomic.      
\begin{VCC}
            b->on = (b->level < 50);
            _(ghost ctrl.d->t = T + 15);
            _(ghost ctrl.deadline = ctrl.d->t)
\end{VCC}
Within the atomic action, we read the water level, and turn on the switch if the level is below the midpoint.
For simplicity, we (unrealistically) assume that we can read the level and update the switch within a single atomic action
(and VCC actually issues a warning about multiple physical accesses within an atomic action). 
In a more realistic driver, we would separate out the read action into a separate action.
\begin{VCC}
            _(bump_volatile_version &ctrl)
\end{VCC}
Because we have updated an owner-approved field of the control protocol (which is owned by this thread), we have to explicitly
indicate to VCC that these fields might change.
\begin{VCC}
        }
    }
    _(unwrap &ctrl)
    _(unwrap &ctrlDeadline)
}
\end{VCC}
Finally, we have to destroy the stack-allocated ghost objects to fultil the proof obligation that every stack-allocated object is destroyed when control leaves the function.
This proof shows that all \vcc{Deadline}s are destroyed.

\section{Using Timers Instead of Deadlines}
Another way to carry out the proof is to use Timers directly in object invariants, in place of \vcc{Deadline}s.
(In this proof, they are essentially interchangeable.) 
In \vcc{BoilerCtrl}, the only change is to change \vcc{d} from a \vcc{PDeadline} to a \vcc{PTimer}.
In the driver, we no longer have to worry about termination:
\begin{VCC}
void boilerDriver(Boiler *b)
    _(maintains \wrapped(b))
    _(writes b)
{
    _(ghost Timer ctrlTimer)
    _(ghost BoilerCtrl ctrl)
    _(ghost {
        ctrl.d = &ctrlTimer;
        ctrl.b = b;
        TimerNew(&ctrlTimer, 15);
        ctrl.t = T+15;
        (&ctrl)->\owns = {ctrl.d, b};
        _(assume b->level >=45 && b->level <= 55)
        _(wrap &ctrl)
    })
    while (1)
        _(writes &ctrl)
        _(invariant ctrl.d == &ctrlTimer && ctrl.b == b)
        _(invariant \wrapped(&ctrl) && ctrl.d->t - T > 10)
    {
        _(assert &ctrl \in \domain(&ctrl) && ctrl.d \in \domain(&ctrl))
        _(atomic &ctrl, ctrl.b, ctrl.d) {
            b->on = (b->level < 50);
            _(ghost ctrl.d->t = T + 15);
            _(ghost ctrl.t = ctrl.d->t)
            _(bump_volatile_version &ctrl)
        }
    }
}
\end{VCC}
It's important to realize that every time we create a timer or change its \vcc{Deadline}, it is essentially an assumption that
it will be destroyed or have its \vcc{Deadline} set again before it expires. This is arguably less natural and harder to understand
than the direct assumptions about time we used for \vcc{Deadline}s. 

\section{Conclusion}
We have shown a general way to extend the VCC methodology to handle timed and hybrid system code.
While the example shown used only a single program thread, the verification is thread-modular, so it is straightforward 
to use the same methodology for multi-threaded code.

The methodology proposed here depends explicitly on the availability of 2-state object invariants, rather than just single-state object invariants. Thus, it cannot be directly used in tools based on separation logic (such as Verifast\cite{Verifast}). However, it might be usable in the framework of separation logic frameworks based on deny-guarantee \cite{DenyGuarantee}.

The two styles of reasoning - one using \vcc{Deadline}s, the other using \vcc{Timer}s - each have their advantages and disadvantages.
Neither methodology is entirely satisfactory, but at this stage, the use of \vcc{Deadline}s seems more practical.
While there is considerable room for improvement, we believe that the methodology is already usable for nontrivial examples.

\bibliography{bibliography}{}

\end{document}